\documentclass[pr,twocolumn,amsmath,amssymb,floatfix]{revtex4-2}
\usepackage{color} 
\usepackage{graphicx}
\usepackage{dcolumn}
\usepackage{bm}
\usepackage{textcomp}
\usepackage{epstopdf}
\usepackage{xcolor}

\usepackage{caption}
\def\D{\mathrm{d}}

\begin{document}

\title{Reducing of the Uncertainty Product of Coherent Light through Multi-Photon Interference \\}

\author{Sangbae Kim$^1$, Joachim St\"{o}hr$^{2,*}$, Fabian Rotermund$^3$, and Byoung S. Ham$^{1,*}$ }

\vspace*{15pt}

\address{
$^1$ School of Electrical Engineering and Computer Science, Gwangju Institute of Science and Technology, Gwangju 61005, Republic of Korea
\\
$^2$ SLAC National Accelerator Laboratory and Department of Photon Science, Stanford University, CA 94035, USA
\\
$^3$ Department of Physics, Korea Advanced Institute of Science and Technology, Daejeon, 34141, Republic of Korea
}

\begin{abstract}
We demonstrate theoretically and experimentally how the diffraction and interferometric resolution limit for single-mode coherent cw laser light can be overcome by multi-photon interference. By use of a Mach-Zehnder interferometer, operated in the single input and single or double output port geometries, we observe a fringe width reduction of the conventional interference pattern, predicted by the wave or single photon quantum theory, by a factor of up to $1/\sqrt{2N}$ through coincident detection of $N=2,3,4$ photons. Our scheme does not require squeezed or entangled light to overcome the standard quantum limit and greatly facilitates precision interferometry experiments. 
\end{abstract}

\maketitle

\section{Introduction}

An important goal and application of modern quantum optics, pioneered by Glauber in the 1960s \cite{glauber:63,glauber:63b,glauber:65}, has been overcoming the limits set by classical wave optics \cite{caves:1981,dowling:2008,abbott:2009,giovannetti:2011,moreau:2019,polino:2020}. By now, quantum optics protocols have led to overcoming  the classical Rayleigh diﬀraction limit in imaging as honored by the 2014 Nobel Prize in Chemistry, proving the existence of the fascinating phenomenon of entanglement honored by the 2022 Nobel Prize in Physics, and creating quantum based metrology methods for the detection of minute distance changes caused by gravitational waves, honored by the 2017 Nobel Prize in Physics. The understanding of the quantum substructure of light in terms of correlated multi-photon quantum states, has also led to paradigm shifts in the broad field of information technology, based on the novel concepts of quantum computation and communication which underlie the development of artificial intelligence. 

Conventional quantum mechanics, such as Dirac's formulation \cite{Dirac-book}, give results that are equivalent to those of the classical wave theory. This fact, expressed by the wave-particle ambiguity, originates from the linearity of conventional quantum mechanics which ignores possible correlations between photons. In quantum theory, interference patterns of light arise through geometric path length differences a \emph{single photon} can take from photon birth to destruction points. The complex electric wave fields that are superimposed  in the classical Huygens-Fresnel principle are simply replaced by a sum over all possible single \emph{photon probability amplitudes}, reflecting the alternative paths a single photon can take \cite{feynman3,liu:2010,stohr-AOP}. While wave interference predicts a continuous macroscopic pattern, the quantum pattern forms probabilistically as a mosaic of single photon detection points (bright spots). 

Modern quantum optics extends Dirac’s original formulation \cite{Dirac-book} of \emph{single} photon or \emph{first order} quantum processes to \emph{multi-photon} processes. The theory may be viewed as the photon-only part of the general theory of light and matter, quantum electrodynamics (QED), which is correct to infinite order. Similar to Feynman's formulation of QED through increasingly higher order correlations between photons and electrons, Glauber's formulation of quantum optics allows for higher order correlations between photons. The behavior of \emph{single} photons corresponds to \emph{first order}, while higher orders are defined through the number of photons, $N >1$, simultaneously involved. These higher order multi-photon correlations are expressed by different multi-photon quantum states of light that can be produced by energetic excitations out of the zero-point quantum vacuum. Some of these multi-photon quantum states have no analogues in a wave-like description and hence give rise to non-classical phenomena.

While the \emph{detection process} is left unspecified in the classical wave formulation, which simply predicts the existence of a continuous macroscopic interference pattern in space, it plays a crucial role in quantum optics. Although the concept or nature of a photon remains controversial among fundamentalists \cite{roychoudhuri:2008,hentschel}, it is the detection process that allows us to state with confidence that we have observed a photon. Photon detection is based on the local destruction of a photon and creation of a photoelectron that gives rise to a detector click. Clicks in different detectors are also used to distinguish whether the detected photons are completely independent or whether there is some kind of correlation between them. Correlations between different photons may be due to simultaneous birth processes in a source or secondary source. Examples are entangled photon pairs generated through spontaneous parametric down conversion \cite{Gerry-Knight} or cloned photons created in stimulated emission \cite{stohr-AOP}.

In quantum optics, the concept of wave interference is extended to the non-trivial concept of photon interference. So-called \emph{multi-photon interference} effects are said to exist when a number of photons detected at points in space and instances in time are deemed \emph{indistinguishable}. Indistinguishability, resulting in interference, exists when two or more photons arrive at single or multiple spatial areas that are smaller than the lateral coherence area (defined by wave-optics), and within a coincident time window that is shorter than the coherence time (defined by the energy bandwidth). When more than single photons are involved, there are more indistinguishable alternatives or photon probability amplitudes that link the photon birth and destruction (detection) processes in space and time. The interference of these multi-photon probability amplitudes change the classical or first order interference pattern. 

Here we investigate multi-photon interference effects for coherent cw laser light. Such light contains a large average number $\langle n \rangle$ of photons, expressed by a collective coherent quantum state or Glauber state of light, and is said to be higher order coherent \cite{loudon}. The photons are statistically distributed in a Poisson probability distribution, which means that the photon arrivals are uncorrelated in time and occur randomly. Coherent laser light most closely resembles a stable classical wave but it has finite quantum uncertainties. 

We discuss how the quantum uncertainties of such light manifest themselves in the fundamental double slit diffraction experiment or a corresponding Mach-Zehnder interference experiment, and how the quantum uncertainties can be reduced through the detection process. We show experimentally that the coincident detection of an increasing number of $N$ photons in coherent light changes the interference pattern, with a strong reduction of the width of the interference fringes with $N$.

\section{The Standard Quantum Limit of Uncertainty}
\label{SS:SQL}

In first order quantum optics, the ultimate precision of a measurement is limited by Heisenberg's uncertainty principle \cite{heisenberg:27}. It was formulated in 1927 through matrix mechanics in terms of the non-commutativity of the position and momentum operators. As discussed in Heisenberg's original paper and in Dirac's \cite{Dirac:27,Dirac:27a} closely related 1927 papers, uncertainty relationships may be phrased more generally in terms of conjugate variables, in analogy to Fourier relationships.  

The position-momentum (e.g. source size and coherent emission angle) uncertainty relation which determines the \emph{diffraction limit} and the energy-time (bandwidth and pulse length) correlation which defines the \emph{transform limit} both have the form $A\,B\geq C$, where $A$ and $B$ represent statistical errors of different conjugate variables and $C$ is a constant \cite{stohr-AOP,stohr:Xrays}. The uncertainty principle only limits the \emph{product} of the two conjugate variables and in principle allows for the trade-off of uncertainties in the two variables by keeping their product constant. 

In the position-momentum or diffraction limit formulation, the trade-off has long been known and is utilized by enlarging the lens of a telescope to get a smaller diffraction limited spot size. By inversion of the optical path, the reduction of a source size, say by an aperture, leads to a concomitant increase of the coherent emission angle. Similarly, in the energy-time or transform limit formulation, one can use a monochromator to reduce the energy bandwidth with a corresponding increase of the coherence time. The inverse method is used to extend the spectral range of lasers into the soft x-ray region through the generation of ultrashort attosecond pulses \cite{schoenlein:2019} which also allow unprecedented temporal resolution, as honored by the 2023 Nobel Prize in Physics. 

Conventional photon counting detectors sum over the number of photons arriving in time. For coherent light, the photon arrival time is random according to a Poisson distribution in time. One therefore observes the same multi-slit diffraction patterns recorded with low-intensity light that is deemed ``coherent'' from a wave point of view and with high-intensity laser light which is higher order coherent \cite{dimitrova-weiss:2008,kim:2023}. Despite the fact that first and higher-order coherent light are formally distinguished in quantum optics \cite{glauber:65,loudon}, for coherent light the diffraction patterns recorded with typical detectors such as position sensitive CCDs are independent of the order of coherence. 

This remarkable fact has the important consequence that x-ray diffraction experiments, where the slits are replaced by periodic 3D arrangements of atoms, give the same Bragg patterns when recorded with conventional detectors, independent of whether the x-ray source is first or higher order coherent! The patterns recorded with conventional x-ray tubes, synchrotron radiation, x-ray free electron lasers \cite{stohr:Xrays} and even phase-stabilized x-ray laser oscillators \cite{zhang:2022} are therefore the same, except for the required recording time.

Starting in the early 1980s, the precision of interferometers were explored for the detection of gravitational wave induced distance changes between widely separated masses \cite{caves:1981}. In this process, the quantum properties of fully coherent laser light emerged as a benchmark for quantum behavior. Such light contains a large average number $\langle n \rangle$ of photons which are statistically distributed in a Poisson probability distribution centered around $\langle n\rangle$, with a standard deviation width of the distribution given by $\sigma=\Delta n= \sqrt{\langle n\rangle}$. This value represents the fundamental ``shot noise'' or ``Poisson noise'' in photon counting. Its quantum mechanical origin lies in the fluctuations in the zero-point quantum vacuum.

The finite quantum uncertainties of single mode laser light are expressed by an uncertainty product that has balanced contributions from the uncertainty in the number of photons $\Delta n$, a quantum property, and in the phase of the associated classical field amplitude, $\Delta \varphi$, according to \cite{dowling:2008}
\begin{eqnarray}
\hspace*{-10pt} \Delta n \Delta \varphi =1~~~\mathrm{where}~~~\Delta n = \sqrt{\langle n\rangle}~~\mathrm{and}~~\Delta \varphi= \frac{1}{\sqrt{\langle n\rangle}}
\label{Eq:laser-uncertainty product} 
\end{eqnarray}
This uncertainty relationship is referred to as the \emph{shot noise limit} or \emph{Standard Quantum Limit} (SQL). It states that the fractional uncertainty in the photon number, $\Delta n/\langle n \rangle$, and the uncertainty in the phase $\Delta \varphi$ of the associated field amplitude both decrease as $1/\sqrt{\langle n \rangle}$ with increasing mean photon number, i.e. both the amplitude and phase of the corresponding electromagnetic wave become better defined. The value of the uncertainty product in (\ref{Eq:laser-uncertainty product}), given here as 1, like the constant $C$ in the Heisenberg uncertainty product, depends on the shape of the distribution functions of the variables and definition of their width \cite{stohr:Xrays}. The signal-to-noise ratio (SNR) in this limit is given by the well-known expression, 
\begin{eqnarray}
\mathrm{SNR}= \frac{\langle n\rangle}{\sqrt{\langle n\rangle}}
\label{Eq:SQL-signal-to-noise} 
\end{eqnarray}

In contrast to the position-momentum and energy-time uncertainty relationships, it is considerably more difficult to trade off between the balanced uncertainties in $\Delta n$ and $\Delta \varphi$ in the SQL expressed by (\ref{Eq:laser-uncertainty product}). Its manipulation requires an extension of the wave and first order quantum descriptions, whose concepts it unites. 

\section{Beyond the Standard Quantum Limit: the Heisenberg Limit}
\label{SS:HL}

The SQL stated by (\ref{Eq:laser-uncertainty product}) corresponds to a transformation of the first order position-momentum principle into a phase space that is spanned by so-called \emph{quadrature operators}, constructed from linear combinations of creation and annihilation operators of the quantized electric field \cite{Gerry-Knight,grynberg:2010}. The four $90^\circ$ phase quadrants completely characterize the complex electric field. 

The reduction of one uncertainty in the SQL at the expense of the other became possible only by considering correlations between photons, which are only allowed in higher order quantum optics. The so-called ``squeezing'' of the uncertainty of one of the variables in the SQL at the expense of the other, was conjectured in 1976 by Yuen \cite{yuen:76} and discussed for interferometry in 1981 by Caves \cite{caves:1981}. It was experimentally demonstrated in 1985 \cite{slusher:1985,andersen:2016}. Squeezing of a coherent state may be visualized by deforming an uncertainty circle in phase space into an uncertainty ellipse of equal area but different lengths or uncertainties along the principal axes \cite{Gerry-Knight,grynberg:2010}. Squeezing is fundamentally a higher order or non-linear concept.

In practice, for precision interferometry one squeezes the more important phase uncertainty at the expense of increasing the number uncertainty. The phase uncertainty in the SQL may then be reduced to the so-called \emph{Heisenberg limit} (HL) according to, 
\begin{eqnarray}
\Delta \varphi_\mathrm{SQL} =  \frac{1}{\sqrt{\langle n\rangle}}~~\longrightarrow~~\Delta \varphi_\mathrm{HL} = \frac{1}{\langle n\rangle}
\label{Eq:HL-phase-reduction}
\end{eqnarray}
Accordance with (\ref{Eq:HL-phase-reduction}), the phase uncertainty in the HL is theoretically reduced relative to the SQL by a factor of $\sqrt{\langle n\rangle}$. For a strong laser beam with a mean photon number of $\langle n\rangle \sim 10^{24}$ as used in gravitational wave detection, this in principle leads to an improvement by an amazing factor of $10^{12}$ \cite{dowling:2008}. Although this improvement is not reached in practice, squeezing has been implemented nevertheless in upgrades of gravitational wave detectors \cite{barsotti:2019}. 

More generally, the HL expresses an extension of the conventional first order Heisenberg uncertainty principle to higher order. This is mathematically defined through photon correlation functions of a well-defined integer number $N$ of photons \cite{scully-zubairy}. In practice, the correlation between the $N$ photons is established through their \emph{coincident detection}. The Heisenberg limit may then be written in the modified form $A\, B\geq C/N$, where $N$ is the number of photons detected in coincidence and $A$ and $B$ are statistical errors of conjugate variables, such as the lateral dimension of a detection area and the planar angle of the observation (detection) cone. It is the coincident detection process that allows the reduction of the width of the measured interference fringes in 1D up to a factor $1/N$, relative to those predicted by the wave theory of light, which corresponds to $N=1$.

In contrast to quadrant squeezing which expresses a continuous interplay of the variables in the SQL by keeping the product constant, the reduction of the width of the interference fringes toward the Heisenberg limit proceeds in discrete steps since $N$ is a positive integer. The most suitable quantum states for reaching the HL are the maximally entangled N00N states of the form $(|N\rangle_a|0\rangle_b+|0\rangle_a |N\rangle_b)/\sqrt{2}$, where $0$ and $N$ are the numbers of photons in the modes $a$ and $b$ \cite{dowling:1998,boto:2000,nagata:2007,duarte2022}. In contrast to the SQL described by (\ref{Eq:laser-uncertainty product}), the HL is expressed by, 
\begin{eqnarray}
\hspace*{-10pt} \Delta n \Delta \varphi =1~~~\mathrm{where}~~~\Delta n = N ~~\mathrm{and}~~\Delta \varphi= \frac{1}{N}
\label{Eq:HL-uncertainty product} 
\end{eqnarray}
The width reduction of the interference pattern was originally demonstrated by use of a beam splitter for the entangled biphoton state $N=2$ \cite{boto:2000} created by spontaneous parametric down conversion \cite{Gerry-Knight}. Similar results were also obtained by use of a double slit geometry \cite{DAngelo:2001}. In both cases, the observed fringe width reduction by a factor of 2 reflects the reduction in phase uncertainty for $N=2$ relative to $N=1$ in (\ref{Eq:HL-uncertainty product}). The more difficult preparation of higher $N$ N00N states has been discussed by Dowling \cite{dowling:2008} and has been implemented for $N=6$ photons which are furthermore hyper-entangled to form a 18-qubit state \cite{wang:2018}.

Our discussion leads to the following simple conceptual distinction of the SQL and the HL, which better reflects their essence than the somewhat confusing names. The ``standard quantum limit'' or SQL is the limit determined by a description of light in first order QED. It describes the behavior of single, independent, and non-interacting photons. This is Heisenberg's original 1927 theory leading to the uncertainty principle \cite{heisenberg:27}, and Dirac's original formulation of what he called ``quantum electrodynamics'' \cite{Dirac:27a}. The name ``Heisenberg limit'' or HL is somewhat of a misnomer since it goes beyond Heisenberg's theory and describes the ultimate limit within Glauber's \emph{multi-photon} quantum optics, which is the photon-only part of the general theory of light and matter, QED, formulated by Feynman, Schwinger and Tomonaga around 1950. The two limits reflect the general principle that the total uncertainty of a measurement has to take into account all factors involved. Thus the consideration of an increasing number of participating photons and their possible correlations allows one to overcome the conventional single photon uncertainty principle. 

In the following we discuss how the classical wave limit can be overcome by use of single mode laser light. Such light, described by a multi-photon collective coherent quantum state, is typically used for the creation of various esoteric quantum states such as number states, squeezed states or entangled states, but we here utilize it directly. Coherent light most closely resembles a classical wave, yet with finite quantum uncertainties. It is this intrinsic quantum behavior, caused by the photon-based graininess, which exemplifies the nature of light beyond the simple wave approximation and its associated limitations. In particular, we show how the conventional interference fringe width of the double slit experiment, performed in a Mach-Zehnder interferometer (MZI) geometry, can be reduced by more than $1/\sqrt{N}$ through coincidence detection of $N$ photons.

\section{Interference Patterns: Double Slit versus MZI Geometry}

\begin{figure}[!h]
\centering
\includegraphics[width=0.35\textwidth]{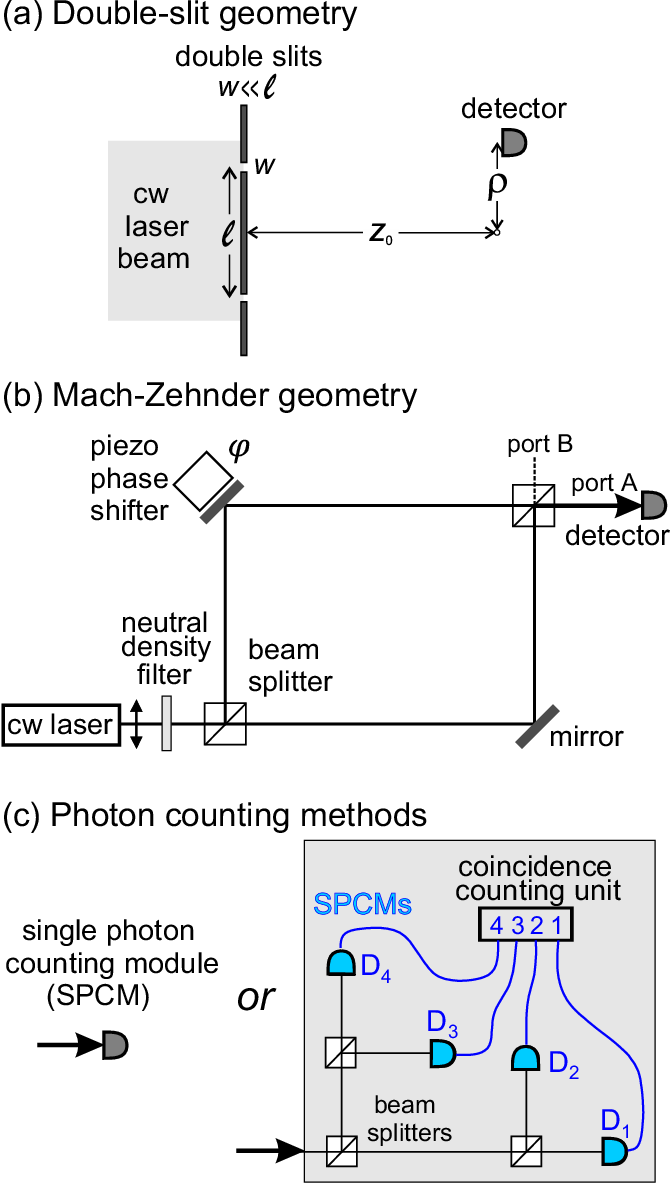}
\caption[]{\small Comparison of double-slit and Mach-Zehnder interferometer (MZI) geometries for the measurement of the multi-photon interference patterns of a fully coherent laser beam. (a) The double slits, separated by a distance $\ell$ much larger than the slit widths $w$, are illuminated by a coherent laser beam. The interference (diffraction) pattern is measured at a large distance $z_0$ by scanning a detectors as a function of distance $\rho$ from the optical axis. (b) Corresponding MZI arrangement yielding identical interference patterns as the double-slit case. The coherent laser beam is injected into one port of a beam splitter, with the other kept open. The relative path length of the light through the two arms of the interferometer are changed by precision movement of one of the mirrors by a piezoelectric transducer. The two beams are combined by a second beam splitter and exit collinearly through one exit port, with the other kept open. (c) Detection scenarios for cases (a) and (b). The single photon interference pattern is measured by use of a single photon counting module (SPCM).  Multi-photon interference patterns of $N$ photons are measured by use of the coincidence detection scheme linking $N$ SPCMs, as shown for $N\leq4$. }
\label{Fig:MZI-geometry}
\end{figure}

In Figs.\,\ref{Fig:MZI-geometry}\,(a) and (b) we illustrate the correspondence of the well-known double slit configuration with the illustrated \emph{specific} input-output paths of a MZI \cite{dimitrova-weiss:2008,kim:2023}.

In Fig.\,\ref{Fig:MZI-geometry}\,(a) the slits are illuminated by a coherent single mode laser beam. We have assumed that the slit separation $\ell$ is much larger than the slit width $w$. The interference or diffraction pattern is recorded at a large distance $z_0$ (Fraunhofer approximation) by scanning a detector as a function of the lateral distance $\rho$ from the optical axis or by use of a position sensitive detector. If the diffracted intensity is calculated classically by use of the Huygens-Fresnel principle of wave interference  or quantum mechanically by adding the number of single photons arriving at a given position $\rho$, one obtains the conventional diffraction pattern given by \cite{stohr:Xrays-16},
\begin{eqnarray}
I^{(1) }(\rho)= \frac{1}{2} \left(1+\cos\left[\frac{k \ell\rho}{ z_0}\right] \right)= \cos^2\left[\frac{k \ell\rho}{2 z_0}\right]
\label{Eq:DS-Single-photon-I} 
\end{eqnarray}
where  $k=2\pi/\lambda$. In (\ref{Eq:DS-Single-photon-I}) we have neglected a $\mathrm{sinc}^2[k w \rho/2 z_0]$ envelope function arising from the finite slit widths $w$, which is nearly unity for $w \ll \ell$. The superscript (1) of the normalized intensity $I$ indicates that the classical pattern corresponds to the single photon, $N=1$, quantum pattern, recorded by averaging over many single photon events by use of a single photon counting module (SPCM) or by use of a position-sensitive detector such as a charge coupled device (CCD), which in each pixel simply adds the charges created by the incident single photons over time.

In the corresponding MZI geometry in Fig.\,\ref{Fig:MZI-geometry}\,(b), utilized in our experiments, a single mode cw laser beam with indicated polarization is injected into one of the two input ports, while the other port contains no photons, i.e. is the quantum mechanical zero-point state. The non-polarizing beam splitter preserves the coherent nature (Poisson statistics) of the input beam \cite{loudon,Gerry-Knight}, and the interference of the beams travelling through the two arms of the MZI is recorded by changing the path length difference $\Delta x$ through the two MZI arms. This is accomplished by moving the mirror in one of the MZI arms by a piezoelectric transducer with sub-wavelength resolution ($\lambda=633$\,nm in our case), as shown. 

The phase shift  $\varphi=k \Delta x =2 \pi \Delta x/\lambda$  between the two MZI paths introduced by the piezoelectric phase shifter leads to an interference pattern when the beams through the two arms are combined by the second beam splitter in the MZI and exit collinearly through the shown exit port A. One may also utilize the unused exit port B, as discussed later. The two schemes in Figs.\,\ref{Fig:MZI-geometry}\,(a) and (b) are equivalent for $\varphi=k \ell\rho/ z_0$, so that the interference pattern is given in analogy to (\ref{Eq:DS-Single-photon-I}) by
\begin{eqnarray}
I^{(1) }(\varphi) = \frac{1}{2} (1+\cos\varphi)= \cos^2 \frac{\varphi}{2}
\label{Eq:Single-photon-MZI} 
\end{eqnarray}
The intensity or number of photons exiting the interferometer are again assumed to be recorded by use of a SPCM that adds charges created by the incident single photons over time. 

In this paper we are interested in the change of the first order quantum pattern (\ref{Eq:Single-photon-MZI}) in higher order quantum optics, corresponding to $N\geq 2$. For either the two-slit or MZI cases, the multi-photon diffraction pattern is then recorded with a detector capable of resolving a specific number of photons $N$ that arrive within a short coincidence window. In practice, such multi-photon coincidence detectors are typically constructed as shown in Fig.\,\ref{Fig:MZI-geometry}\,(c) through a combination of 50/50 beam splitters and SPCMs linked by coincidence circuits. We will use this detection scheme to compare the interference patterns for $1\leq N \leq 4$ photons.

For convenience, our experimental results are obtained by use of the MZI scheme in Fig.\,\ref{Fig:MZI-geometry}\,(b), but the double slit configuration would yield the same patterns \cite{dimitrova-weiss:2008,kim:2023}. In order to avoid false coincidences within a typical detection time interval ($\sim$\,1\,ns), the incident laser beam can be attenuated by a neutral density filter as shown, since it maintains the Poisson statistics of the unattenuated coherent laser beam \cite{kim:2023}. We will later show, that the same results may be conveniently obtained with higher intensity beams and modified detection as discussed in Sect.\,\ref{SS:alternative-detection}. Experimental details are given in the ``Methods'' section \ref{S:Methods} below.

\section{From Single to Multi-Photon Interference Patterns}

\subsection{Experimental Results}

\begin{figure}[!h]
\centering
\includegraphics[width=0.35\textwidth]{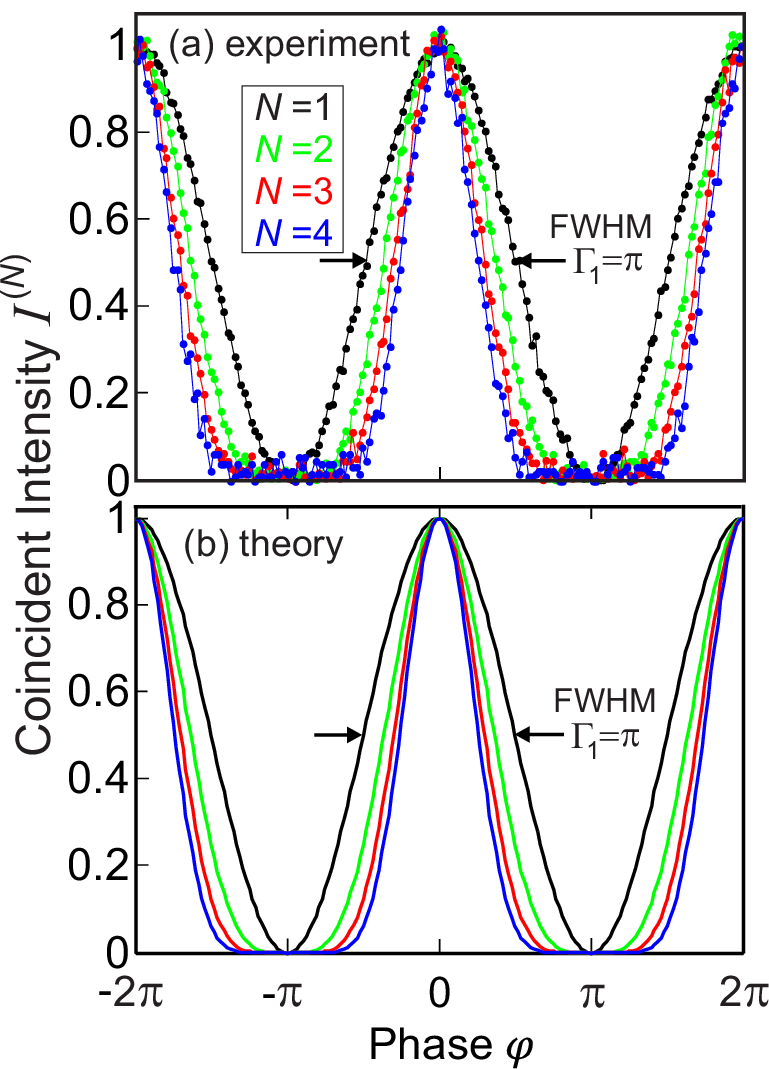}
\caption[]{(a) Measured intensity as a function of the number $N$ of photons detected in coincidence. The classical or $N=1$ finge width (FWHM) is $\Gamma_1=\pi$.(b) Theory according to (\ref{Eq:N-photon-I}). The narrowing of the FWHM of the central fringe, $\Gamma_N$, with $N$, relative to that of the classical width, $\Gamma_1=\pi$, of the black curve, is given by (\ref{Eq:FWHM-N}) below. }
\label{Fig:Coinc-intensity}
\end{figure}

The measured interference pattern for the conventional single photon, $N=1$, case representing also the classical wave pattern (\ref{Eq:Single-photon-MZI}), is shown in black in Fig.\,\ref{Fig:Coinc-intensity}\,(a). It is compared to the intensity normalized patterns for $N=2,3,4$ recorded in coincidence. 

Coincidence detection with 1-photon detectors illustrated in Fig.\,\ref{Fig:MZI-geometry}\,(c) typically requires attenuating the incident laser beam to avoid accidental coincidences. This was fulfilled for the data shown in Fig.\,\ref{Fig:Coinc-intensity}\,(a) recorded with an incident rate of $R =2\times 10^7$\,photons/sec and a detection time window $\tau \simeq 6$\,ns, yielding a single photon arrival probability of $R\,\tau \simeq 0.12$.

The FWHM width of the central fringe, $\Gamma_N$, in Fig.\,\ref{Fig:Coinc-intensity}\,(a) reduces with $N$ according to $\Gamma_1= 3.14=\pi$, $\Gamma_2=2.28$, $\Gamma_3= 1.88$, and $\Gamma_4= 1.64$. The maximum photon count rate for $N\!=\!1$ of $2\times 10^7$ per second drastically decreased with increasing $N$, resulting in $\simeq 5.2 \times 10^5$ ($N\!=\!2$), $\simeq 1.7\times 10^4$ ($N\!=\!3$), and $\simeq 1.3 \times 10^3$ ($N\!=\!4$) counts per second. 

\subsection{Quantum Optics Theory}

The coherent laser beam injected into the single port of the MZI corresponds to a single mode  \emph{collective coherent state} $|\alpha\rangle$  given by \cite{loudon}
\begin{equation}
|\alpha \rangle = \mathrm{e}^{-\frac{|\alpha|^2}{2}}
\sum_{N=0}^\infty \frac{\alpha^N}{\sqrt {N! \,} }\, |N \rangle
 \label{Eq:SM-coh-state}
\end{equation}
The state contains a mean photon number $\langle N \rangle =|\alpha|^2$ and is normalized according to $\langle \alpha|\alpha \rangle= 1$. The other input port is left open, corresponding to the vacuum state $|0\rangle$. The beam splitter (BS), assumed to be lossless, changes the incident state $|\alpha \rangle, |0 \rangle$ into two equal parts containing half of the incident number of photons, each. The output beams emerge in the reflected (r) and transmitted (t) directions relative to the input direction of $|\alpha\rangle$ according to the input-output relation \cite{loudon,Gerry-Knight,kim:2009},
\begin{eqnarray}
|\alpha \rangle, |0 \rangle \, \stackrel{\mathrm{BS}\,}{\Longrightarrow}  \mathrm{i} \left|\frac{ \alpha}{\sqrt{2}}\right \rangle_\mathrm{r}, \, \left|\frac{ \alpha}{\sqrt{2}}\right \rangle_\mathrm{t}
\label{Eq:coherent-state-out}
\end{eqnarray}
The output beams emerging from the first beam splitter of the MZI are again coherent states, with the subscripts indicating the two propagation directions or modes through the two branches of the MZI. The factor i in (\ref{Eq:coherent-state-out}) indicates that the reflected beam experiences a phase shift of $90^\circ$, while the transmitted beam is not phase shifted \cite{degiorgio:1980}. This phase convention is unimportant for the measured intensities.

The output states of the two exit ports A and B of the MZI defined in Fig.\,\ref{Fig:MZI-geometry}\,(b) are the superimposed beams through the two arms. The complete action of the MZI including the phase shifter may be derived as \cite{Mandel:87b,shin:1999},
\begin{eqnarray}
|\alpha\rangle , |0 \rangle\, \stackrel{ \mathrm{MZI}\,}{\Longrightarrow}  \frac{\mathrm{i}(1+ \mathrm{e}^{\mathrm{i} \varphi})}{2}   |\alpha \rangle_\mathrm{A},\frac{(1- \mathrm{e}^{\mathrm{i} \varphi})}{2} |\alpha \rangle_\mathrm{B}
\label{Eq:final-MZI-coh-state}
\end{eqnarray}

For a loss-less MZI, the average number of detected photons emerging from the two output ports  A and B conserves the number of injected photons $\langle N \rangle_\mathrm{in}= |\alpha|^2$ according to,
\begin{eqnarray}
 \langle N \rangle_\mathrm{A} &=&  |\alpha|^2 \cos^2\frac{\varphi}{2}
 \nonumber \\
  \langle N \rangle_\mathrm{B} &=& |\alpha|^2 \sin^2\frac{\varphi}{2}
 \label{Eq:MZI-one-port-out-number} 
\end{eqnarray}
For a path (phase) balanced MZI, $\varphi=0$, all photons emerge from port A while port B is ``dark'' due to destructive interference.

The single-mode collective coherent state defined by (\ref{Eq:SM-coh-state}) is a sum of number states $|N\rangle$ which contribute to the Poisson probability distribution around the mean value $\langle N \rangle$ with weight factors given by 
\begin{equation}
P_N = \mathrm{e}^{-|\alpha|^2} \frac{|\alpha|^{2 N}}{N!} = \mathrm{e}^{-\langle N\rangle} \frac{\langle N\rangle^{N}}{N!}
 \label{Eq:PN-coh-state}
\end{equation}
It may be rewritten to express the arrival probability of $N$ photons within a given detection time interval of length $\tau$. Denoting the average count rate as $R=\langle N\rangle/\langle t\rangle $, where $\langle t\rangle$ is a sufficiently long time interval to obtain a reliable average, the $N$ photon coincident detection probability within a detection time window $\tau$ is given by,
\begin{equation}
P_N(R,\tau) =\mathrm{e}^{- R \,\tau}\,\frac{(R \,\tau)^N } { N! }
\end{equation}
This arrival probability allows us to express the $N$ photon coincidence count rate in terms of the single photon count rate. The ratio of $N$- versus 1-photon counts within an interval $\tau$ is obtained as, 
\begin{equation}
R_{N,1}=\frac{P_N}{P_1} =\frac{(R \,\tau)^{N-1} } { N!}
\label{Eq:N-vs-1-counts} 
\end{equation}
Assuming identical detection time intervals of $\tau \simeq 6$\,ns, the experimental count ratios $R_{2,1} = 5.2\times 10^{-2}, R_{3,1} = 1.6\times 10^{-3}, R_{4,1} = 1.4\times 10^{-4}$ compare to the theoretical ones $R_{2,1} = 3\times 10^{-2}, R_{3,1} = 6\times 10^{-4}, R_{4,1} = 9\times 10^{-6}$. The large discrepancy for $N=4$ is a consequence of the low 4-photon  coincident count rate relative to the dark count rate of 50 per second. 

The relation (\ref{Eq:N-vs-1-counts}) also reveals why multi-photon coincidence detection is required to change the conventional 1-photon interference pattern as illustrated in Fig.\,\ref{Fig:Coinc-intensity}\,(a). For a coherent beam, the single-photon rate exceeds the multi-photon coincidence rates by orders of magnitude. One therefore needs to pick out the much lower joint $N$-photon counts  through the coincidence trick \cite{ou:2006}. This corresponds to the quantum projection of the collective coherent state $|\alpha\rangle$ given by (\ref{Eq:SM-coh-state}) onto its specific number substates $|N\rangle$. It is this quantum projection method which is at the heart of picking apart the near-classical wave behavior of the collective state $|\alpha\rangle$.

For our case of the beam emerging from port A in Fig.\,\ref{Fig:MZI-geometry}\,(b), the change of the one photon interference pattern $I^{(1) }(\varphi)$ given by (\ref{Eq:Single-photon-MZI}) to the $N$ photon pattern follows from the special properties of the coherent Glauber state. For a coherent state, the $N^{th}$-order photon-photon correlation function factors into the product of $N$ single photon correlation functions \cite{glauber-titulaer:65,loudon,stohr:Xrays}. As a consequence, the $N$-photon coincidence interference pattern is simply the first order pattern to the $N^{th}$ power, 
\begin{eqnarray}
I^{(N) } (\varphi) = \left[\cos^2 \frac{\varphi}{2}\right]^N
\label{Eq:N-photon-I} 
\end{eqnarray}
This simple relationship reflects the fact that the $N$ photons within the detection window $\tau$ are indistinguishable,  and hence may interfere. Unlike the classical interference of complex wave-fields $E$ which possess a phase, the quantum mechanical ``interference'' of phase-less photons arises more specifically from interference of their path-related probability amplitudes. Multi-photon interference is a non-linear process which becomes allowed when Dirac's conventional first order quantum mechanics is extended to higher order.

The theoretically predicted interference pattern (\ref{Eq:N-photon-I}) for $1\leq N \leq 4$ is shown in Fig.\,\ref{Fig:Coinc-intensity}\,(b) and it fully agrees with the experimentally observed pattern in Fig.\,\ref{Fig:Coinc-intensity}\,(a). Before we discuss the scaling of the fringe width with $N$ which is of particular interest, we briefly discuss another detection scheme that yields the same results. 

\subsection{Alternative Detection Scheme}
\label{SS:alternative-detection}

\begin{figure}[!h]
\centering
\includegraphics[width=0.35\textwidth]{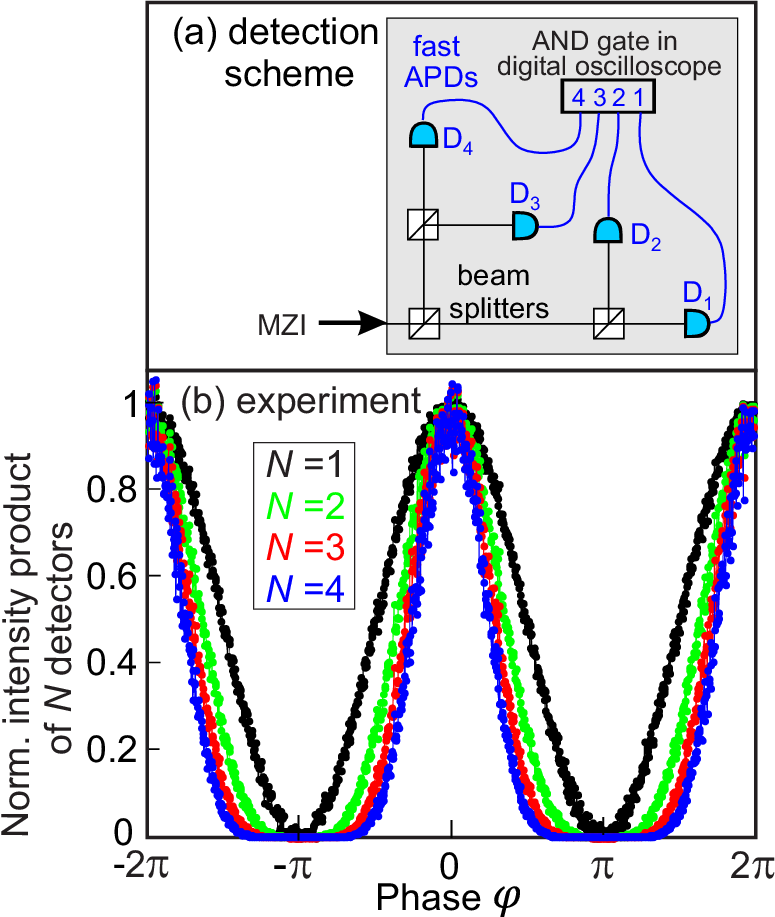}
\caption[]{\small (a) Modified detection scheme of Fig.\,\ref{Fig:MZI-geometry}\,(c). The single photon counting modules (SPCMs) are replaced by avalanche photo diodes (APDs), and the coincidence signal is calculated by a built-in AND gate function of a fast digital oscilloscope. (b) Experimental interference patterns. }
\label{Fig:CW-detection}
\end{figure}

For a coherent cw laser beam, the simple quantum optics evolution of the interference pattern with increasing order $N$ expressed by (\ref{Eq:N-photon-I}) allows us to conveniently simplify the detection process, as shown in Fig.\,\ref{Fig:CW-detection}\,(a). The single-photon counting modules (SPCMs) are replaced by fast avalanche photo diodes (APDs) and their intensity product is calculated by use of a built-in AND gate function of a fast digital oscilloscope (see Sect.\ref{S:Methods} below). The so-measured interference patterns are shown in Fig.\,\ref{Fig:CW-detection}\,(b). They are identical within statistics to the coincident patterns in  Fig.\,\ref{Fig:Coinc-intensity}\,(a).

In Figs.\,\ref{Fig:Coinc-intensity}\,(a) and \ref{Fig:CW-detection}\,(b), the scalability with $N$ is in principle up to the total photon number of the input light. In practice, a linear detector response is required which in our case limited the incident intensity to $\simeq 10^{13}$\,photons/s (see Sect.\,\ref{S:Methods}).  The use of high intensity coherent beams is a great benefit of the present method in comparison to quantum sensing based on N00N states, whose production becomes increasingly difficult for $N>2$ \cite{nagata:2007,dowling:2008}.

\subsection{Fringe Width Reduction through Multi-Photon Interference}

The reduction of the full-width-at-half-maximum (FWHM) of the fringe width with $N$, schematically indicated in Fig.\,\ref{Fig:Coinc-intensity}, represents to what degree the conventional diffraction limit can be overcome. The interference fringes narrow around $\varphi=0$ and the equivalent phase angle $\varphi=\pm 2\pi$ at the expense of intensity loss around $\varphi=\pm \pi$. The shown intensity distributions are normalized to their peak value to better reveal the fringe narrowing. 

\begin{figure}[!h]
\centering
\includegraphics[width=0.35\textwidth]{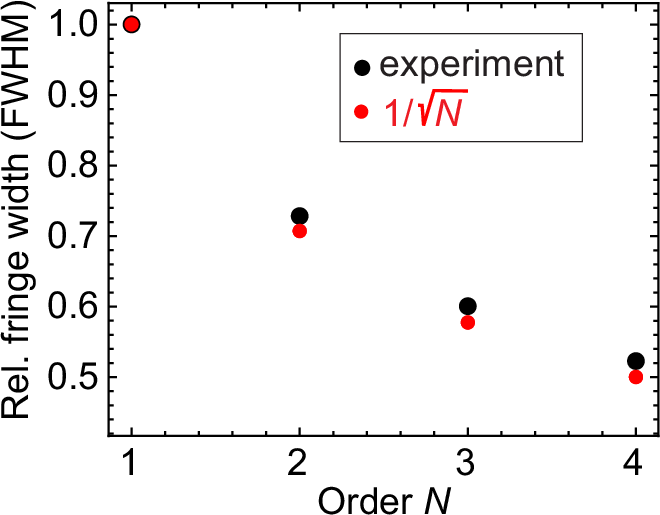}
\caption[]{Relative reduction of the fringe width (FWHM) of the patterns in Fig.\,\ref{Fig:Coinc-intensity} with order $N$,  according  (\ref{Eq:FWHM-N}) (black) and scaling $1/\sqrt{N}$ (red). }
\label{Fig:Fringe-width}
\end{figure}

The FWHM of the central fringe of the pattern (\ref{Eq:N-photon-I}), $\Gamma_N$ (see Fig.\,\ref{Fig:Coinc-intensity}), reduces  as a function of $N$ relative to the first order $N=1$ width according to, 
\begin{eqnarray}
\frac{\Gamma_N}{\Gamma_1}=\frac{4}{\pi} \mathrm{arcsec}\!\left[2^{\frac{1}{2N}}\right]\simeq \frac{1}{\sqrt{N}}
\label{Eq:FWHM-N} 
\end{eqnarray}
The reduction of the fringe width (FWHM) with $N$ relative to the classical width according to (\ref{Eq:FWHM-N}) is shown in Fig.\,\ref{Fig:Fringe-width} for $N\leq 4$ (black points). Also shown is a reduction according to $1/\sqrt{N}$ (red points), which is a good approximation. 

The slight deviation of the black data points from the $1/\sqrt{N}$ dependence shown in red in Fig.\,\ref{Fig:Fringe-width} is due to the fact that the minimum possible quantum mechanical uncertainty product holds only for Gaussian distributions \cite{kennard:1927}. Given a single photon intensity distribution $I(x)$ with a peak value $I(0)=1$ in the detector plane, the FWHM of the normalized distributions $I(x)$ and $[I(x)]^N$ scale with $1/\sqrt{N}$ only for Gaussian distributions $I(x) = \mathrm{e}^{-a^2 x^2/2\Gamma^2}$, where $\Gamma= a \sigma$ is the FWHM, $a = 2\sqrt{\ln{4}}$ and $\sigma$ is the rms width. In this case we have the exact scaling law \cite{stohr-AOP}
\begin{eqnarray}
\int_{-\infty}^\infty [I(x)]^N \,\D x= \frac{1}{\sqrt{N}} \int_{-\infty}^\infty  I(x)\, \D x
\label{Eq:effective-width-relation}
 \end{eqnarray}
It is only approximately fulfilled for our case of $I(x) = \cos^2 (x)$ given by (\ref{Eq:Single-photon-MZI}) and $[I(x)]^N = [\cos^2 (x)]^N$ expressed by (\ref{Eq:N-photon-I}) when integrated over a cycle interval $-\pi$ to $\pi$.

\section{Further Fringe Width Reduction by Correlation of both MZI Output Ports}

\begin{figure}[!h]
\centering
\includegraphics[width=0.4\textwidth]{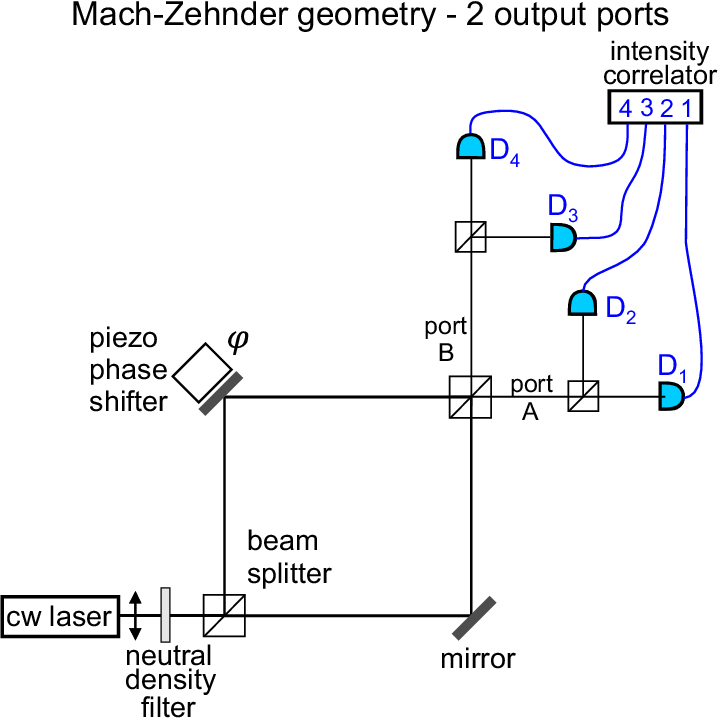}
\caption[]{Alternative MZI geometry for the measurement of the multi-photon cross interference patterns between the two output ports A and B. Instead of a single beam output shown in Figs.\,\ref{Fig:MZI-geometry} and \ref{Fig:CW-detection}, both exit beams are used to feed two single-photon coincidence detectors D$_1$ and D$_2$ in one and  D$_3$ and D$_4$ in the other output arm.  Multi-photon interference patterns of 4 photons can now be measured as shown. }
\label{Fig:Alternative-MZI-geometry}
\end{figure}

The multi-photon interference fringe width can be further reduced by correlating the intensities (or photons) emerging from the two MZI output ports A and B, as illustrated in Fig.\,\ref{Fig:Alternative-MZI-geometry} for the case of correlating the 2-photon coincidence patterns of A and B, respectively. 

\begin{figure}[!h]
\centering
\includegraphics[width=0.4\textwidth]{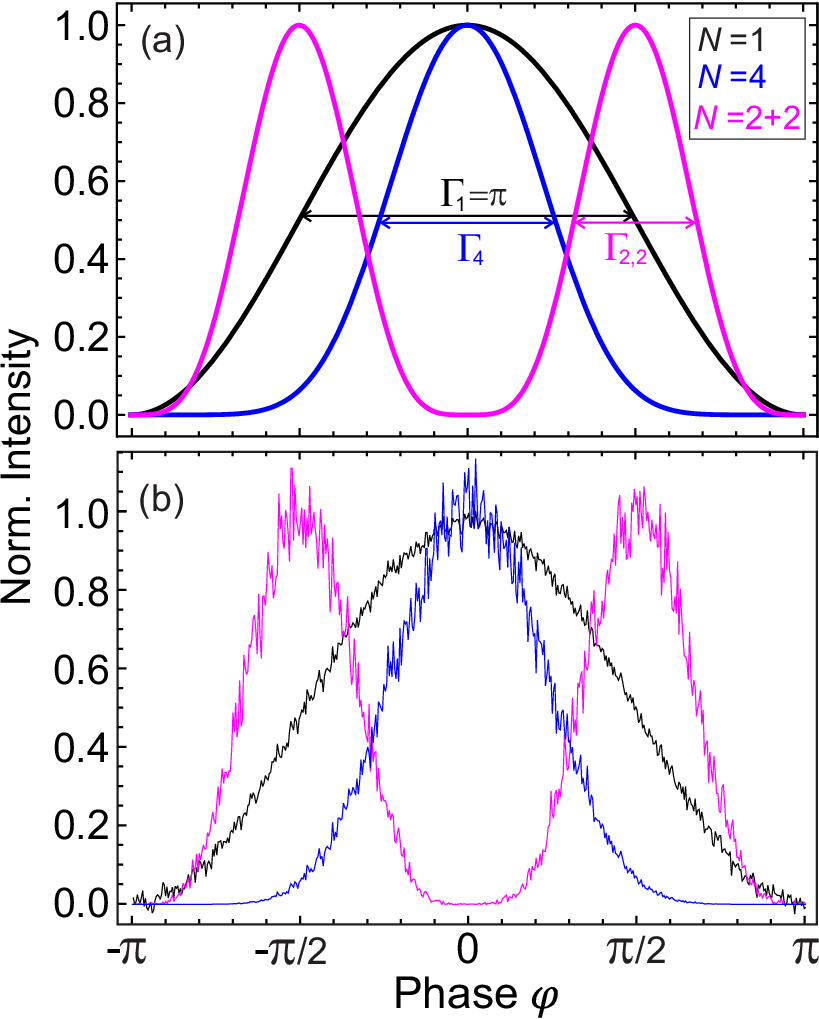}
\caption[]{(a) Port A interference pattern for $N=1$ (black, classical pattern) and $N=4$ coincidence pattern (blue) according to (\ref{Eq:N-photon-I}), in comparison to the cross correlation pattern of ports A and B for $N=4$ (2 photons in each port) according to (\ref{Eq:N-photon-cross-intensity}) (magenta). The fringe width of the magenta is reduced relative to the blue pattern by a factor of $1/\sqrt{2}$. \newline
(b) Corresponding experimental results. The black and blue curves are taken from Fig.\,\ref{Fig:CW-detection}\,(b). The magenta curve was  recorded with the same cw-like detection scheme of Fig.\,\ref{Fig:CW-detection}\,(a), but the $N=2$ port A intensity was correlated with the $N=2$ port B intensity as shown in Fig.\,\ref{Fig:Alternative-MZI-geometry}. }
\label{Fig:single-vs-two-ports}
\end{figure}

According to (\ref{Eq:MZI-one-port-out-number}) the interference patterns for the two output ports A and B are out of phase by $\pi$. If we correlate the intensities  of ports A and B, both composed of half of the incident number $N$ as shown in  Fig.\,\ref{Fig:Alternative-MZI-geometry}, the single output port A intensity for $N$-photon given by (\ref{Eq:N-photon-I}), is replaced by the cross-correlated intensity of $N/2$ photons per port according to,
\begin{eqnarray}
I^{(N) }_\mathrm{cross}  = I^{\left(\!\frac{N}{2}\!\right ) }_\mathrm{A} \,I^{\left(\!\frac{N}{2}\!\right) }_\mathrm{B}= \left[\cos^2 \frac{\varphi}{2}\right]^{\frac{N}{2}} \left[\sin^2 \frac{\varphi}{2}\right]^{\frac{N}{2}} 
\label{Eq:N-photon-cross-intensity} 
\end{eqnarray}
This theoretical 2-port pattern for $N=4$ (2 photons in each port) is shown as a magenta curve in Fig.\,\ref{Fig:single-vs-two-ports}\,(a) together with the classical or $N=1$ reference pattern (black) given by (\ref{Eq:Single-photon-MZI}) and the $N=4$ single output-port pattern (blue) expressed by (\ref{Eq:N-photon-I}).

The single-port classical fringe width $\Gamma_1= 3.14=\pi$ reduces to the single-port $N=4$ coincidence width $\Gamma_4= 1.64$ as previously shown in Fig.\,\ref{Fig:Coinc-intensity} and plotted in Fig.\,\ref{Fig:Fringe-width}. Denoting the fringe width of the two-port correlated $N$ photon pattern given by (\ref{Eq:N-photon-cross-intensity}) as $\Gamma_{\frac{N}{2},\frac{N}{2}}$, it can be shown that for the same $N$, we have,
\begin{eqnarray}
\Gamma_{\frac{N}{2},\frac{N}{2}}= \frac{\Gamma_N}{\sqrt{2}}
\label{Eq:Fringe-width-single-double} 
\end{eqnarray}
Therefore we obtain an additional $1/\sqrt{2}$ fringe width reduction for the two-port relative to the one-port case. This is confirmed by the experimental results shown in Fig.\,\ref{Fig:single-vs-two-ports}\,(b). The black and blue curves are those in Fig.\,\ref{Fig:CW-detection}\,(b). The magenta curve was also recorded with the APDs and built-in AND gate function of a fast digital oscilloscope shown in Fig.\,\ref{Fig:CW-detection}\,(a), but the $N=2$ port A intensity was correlated with the $N=2$ port B intensity as shown in Fig.\,\ref{Fig:Alternative-MZI-geometry}.

\section{Discussion and Conclusions}

We have theoretically and experimentally investigated interference effects and the associated minimum fringe width observed for single-mode coherent cw laser light. Such light most closely resembles a classical electromagnetic wave of well defined amplitude and phase. This is reflected by the fact that both Young's fundamental double slit experiment and related interferometer-based metrology schemes carried out with laser light and conventional detection schemes reproduce the interference patterns predicted by the classical wave-based Huygens-Fresnel principle. 

Despite the perceived wave-like behavior of coherent laser light, its fundamental description through the photon concept in QED is more intricate and complex. As discussed in this paper, the quantum complexity may be utilized to overcome classical limitations. The classical paradigms are shown to hold only in first order QED, as reflected by the wave-particle ambiguity supported by conventional quantum mechanics. The equivalence of the classical wave and the first order QED description is evidenced by the same predicted diffraction limit. In first order QED or conventional quantum mechanics, measurement uncertainties are determined by Heisenberg's position/momentum uncertainty principle. This principle, in the form of the corresponding area/solid-angle relation underlies the diffraction limit \cite{stohr-AOP}. It also emerges from the classical Huygens-Fresnel principle, as already shown in the late 1800s by Abbe \cite{abbe:1873} and Lord Rayleigh \cite{rayleigh:1897}.

We show, that the limitations of the wave theory or the equivalent first order quantum theory, can be overcome through the quantum optics formulation of higher order coherence, which is accompanied by the concept of multi-photon interference. While finite quantum uncertainties are always present due to the fluctuations in the zero-point quantum vacuum, their relative size can be increasingly reduced in higher order leading to the Heisenberg limit. We show that for coherent light the reduction of the fringe width according to the $1/N$ Heisenberg limit is not reached, yet a substantial reduction according to $1/\sqrt{2N}$ is accomplished with great experimental ease. This exceeds the previously predicted $1/\sqrt{N}$ optimum scaling \cite{dowling:1998}. 

In principle, the use of entangled N00N states with $N>2$ may yield the $1/N$ fringe width reduction according to the HL. In practice, it is difficult to prepare such states for $N>2$ \cite{ou:2006,nagata:2007,dowling:2008}. Another issue in quantum sensing with either N00N or  squeezed states is photon loss of the weak beams in the transmission process. In this context, the demonstration of the direct use of high intensity laser light in conjunction with the alternative intensity-product sensing discussed in Sect.\,\ref{SS:alternative-detection} provides a great practical benefit.

We note that for the two-dimensional case of a coherent circular source, the 2D area of the central Airy disc which defines the diffraction limit of a telescope or microscope, indeed scales as $1/N$ with the number of photons $N$ detected in coincidence at a given point or at two points that are symmetrically separated from the optical axis \cite{stohr-AOP}. For optical imaging with a 1\,mW laser ($\sim 10^{15}$ photons per second), a 2D photo-detector array with $10^7$ pixels, and an overall 1\% instrumental efficiency, each pixel would still accumulate on average $10^6$ photons per second. If the intensity of all pixels is multiplied using the detection scheme in Fig.\,\ref{Fig:CW-detection}, one would obtain a thousand times-enhanced resolution with sufficient SNR. 

Finally, we would like to point out that the interference theory for the MZI case is considerably simpler than for the double slit case. The MZI case can essentially be described by a single mode theory, since the modes exiting the two interferometer ports are orthogonal. In contrast, the quantum mechanical theory of the double slit case is considerably more complicated, both in the single- and multi-photon treatment \cite{stohr:Xrays-16}. It requires a two-mode theory since interference paths from the two slits to a detection point involve finite angles between possible $\vec k$ directions. The double slit case, however, more clearly reveals a new paradigm emerging from the quantum treatment of diffraction. It states that the observed diffraction patterns are direct signatures of the quantum states of light emitted by the slits. In practice, different quantum states of light may be created at the slit positions by modification of a single mode laser beam \cite{stohr:Xrays-16}.

\section{Methods}
\label{S:Methods}

The Mach-Zehnder interferometer (MZI) comprized two 50/50 non-polarizing beam splitters (BS). The wavelength of the cw laser (Thorlabs HNL020L) was 633\,nm with a coherence time of 200\,ns. The incident polarization direction is in the MZI plane as shown in Fig.\,\ref{Fig:MZI-geometry}\,(b). Four single-photon counting modules (SPCMs) (Excelitas SPCM-AQRH-15) with a dynamic range  $> 3.5\times 10^7$ photons/s, a dark count rate of 50 photons/s, and a temporal resolving time of $\sim\,350$\,ps generated the intensities utilized in the coincidence measurements. For the  intensity correlations of $N=1-4$ photons, a four-channel coincidence counting unit (Altera, DE2) post-selected a specific number $N$ photons from the Poisson distribution of the coherent quantum state. 

For the experimental results in Fig.\,\ref{Fig:CW-detection}, the input laser power before entering the MZI was set at 3\,$\mu$W ($\simeq 10^{13}$\,photons/s) and fast silicon avalanche photodiodes (APDs) (Thorlabs, APD-110A), each with a linear response up to $1.5\,\mu$W, replaced the SPCMs in Fig.\,\ref{Fig:MZI-geometry}\,(c). Also, the coincidence counting unit was replaced by a four-channel digital oscilloscope (Yokogawa, DL9040) with a maximum 5\,GHz sampling rate and 500\,MHz bandwidth. Each data point was averaged 30 times. The intensity products of the APDs were calculated by use of a built-in AND gate function of the oscilloscope. For both detection schemes used for Figs.\,\ref{Fig:Coinc-intensity}\,(a) and \ref{Fig:CW-detection}\,(b), the piezoelectric phase shifter in Fig.\,\ref{Fig:MZI-geometry}\,(b) was scanned for 36 seconds in a forward mode from $\varphi=-2\pi$ to $\varphi=2\pi$ at steps of $2\pi/180$. Thus, there are a total of 360 data points in each curve. The mean photon number of each data point was counted for 0.1 second.
\newline\newline
\emph{Acknowledgements}
\newline
JS would like to thank Robin Santra and Jianbin Liu for valuable discussions.
\newline
\emph{Funding}
\newline
This research was supported by the MSIT (Ministry of Science and ICT), Korea, under the
ITRC (Information Technology Research Center) support program (IITP 2024-2021-0-01810) supervised
by the IITP (Institute for Information and Communications Technology Planning and Evaluation). 
\newline
\emph{Author contribution}
\newline
JS proposed the experiment and together with BSH wrote the paper. BSH supervised the experiments which were conducted by SK, and analyzed the data. FK contributed to the discussion. 
\newline
\emph{Competing Interests}
\newline
The authors declare no competing interest.
\newline
\emph{Data Availability}
\newline
All data generated in this study are available on request from BSH.

\end{document}